\documentclass[pre, aps,twocolumn,floatfix,superscriptaddress]{revtex4-1}
\usepackage{amsmath,amssymb,graphicx,cases}
\usepackage{epsfig}
\usepackage{textcomp}
\usepackage{epstopdf}
\usepackage{float}
\usepackage{braket}
\usepackage[normalem]{ulem}

\usepackage[normalem]{ulem}
\newcommand{\stkout}[1]{\ifmmode\text{\sout{\ensuremath{#1}}}\else\sout{#1}\fi}

\newcommand{\nn}{\nonumber}
\newcommand{\be}{\begin{equation}}
\newcommand{\ee}{\end{equation}}
\newcommand{\bea}{\begin{eqnarray}}
\newcommand{\eea}{\end{eqnarray}}

\usepackage[colorlinks=true, urlcolor=blue, anchorcolor=blue, citecolor=blue,filecolor=blue,linkcolor=blue,menucolor=blue]{hyperref}

\begin{document}
\title{Anomalous scaling and first-order dynamical phase transition in large deviations of the Ornstein-Uhlenbeck process}

\author{Naftali R. Smith}
\email{naftalismith@gmail.com}
\affiliation{CNRS-Laboratoire de Physique Th\'eorique de l'Ecole Normale Sup\'erieure, 24 rue Lhomond, 75231 Paris Cedex, France}
\affiliation{Department of Solar Energy and Environmental Physics, Blaustein Institutes for Desert Research, Ben-Gurion University of the Negev,
Sede Boqer Campus, 8499000, Israel}


\begin{abstract}

We study the full distribution of $A=\int_{0}^{T}x^{n}\left(t\right)dt$, $n=1,2,\dots$, where $x\left(t\right)$ is an Ornstein-Uhlenbeck process. We find that for $n>2$ the long-time ($T \to \infty$) scaling form of the distribution is of the anomalous form $P\left(A;T\right)\sim e^{-T^{\mu}f_{n}\left(\Delta A/T^{\nu}\right)}$
where $\Delta A$ is the difference between $A$ and its mean value, and the anomalous exponents are $\mu=2/\left(2n-2\right)$, and $\nu=n/\left(2n-2\right)$.
The rate function $f_n\left(y\right)$, that we calculate exactly, exhibits a first-order dynamical phase transition which separates between a homogeneous phase that describes the Gaussian distribution of typical fluctuations, and a ``condensed'' phase that describes the tails of the distribution.
We also calculate the most likely realizations of $\mathcal{A}(t)=\int_{0}^{t}x^{n}\left(s\right)ds$
and the distribution of $x(t)$ at an intermediate time $t$ conditioned on a given value of $A$.
Extensions and implications to other continuous-time systems are discussed.

\end{abstract}

\maketitle

\section{Introduction}
The study of fluctuations in stochastic systems is of central importance in non-equilibrium statistical mechanics and probability theory. Rare events, or large deviations, are of particular interest 
 \cite{Varadhan,O1989,DZ,Hollander,Bray, Majumdar2007, T2009,Derrida11, RednerMeerson, Vivo2015, MeersonAssaf2017, MS2017,Touchette2018}. A standard way to characterize fluctuations in dynamical systems is by considering distributions 
of ``dynamical observables'' which are given by integrating the stochastic process over time,
\be
A =\int_{0}^{T}u\left(X\left(t\right),\dot{X}\left(t\right)\right)dt\,,
\ee
where $X(t)$ is the stochastic process and $u\left(X, \dot{X}\right)$ is an arbitrary function. 
For ergodic systems, in the long-time limit $T \to \infty$ the time average $A/T$ converges to its ensemble-average value, but the fluctuations from this value depend on the temporal correlations of the process, and can exhibit non-equilibrium features even if the system is in equilibrium.
A general theory for the study of such fluctuations was developed, based on the Feynman-Kac formula \cite{Bray, Majumdar2007, DonskerVaradhan,Ellis,T2009,Touchette2018}. This theory is sometimes referred to as the Donsker-Varadhan (DV) theory. 
Under quite general conditions, capturing a broad class of physical systems, the theory predicts that at long times the probability density function (PDF) $P\left(A;T\right)$ of $A$ obeys a large-deviation principle,
\be
\label{eq:DVscaling}
P\left(A;T\right)\sim e^{-TI\left(A/T\right)}, \quad T \to \infty,
\ee 
i.e., the limit $-\lim_{T\to\infty}\ln P\left(aT^{\nu};T\right)/T^{\mu}=I\left(a\right),$
with the standard exponents $\mu=\nu=1$ exists with a
``rate function'' $I(a)$ that is nonnegative, convex, and vanishes when its argument $a = A/T$ equals its corresponding ensemble-average value.
Thus, $I(a)$ quantifies large deviations of $a$ which become exponentially unlikely in $T$.
The rate function is calculated from the dominant eigenvalue of the Feynman-Kac equation for the generating function of $A$.
The calculation boils down to solving an auxiliary problem of finding the largest eigenvalue of a ``tilted operator'', and then applying a Legendre-Fenchel transform to the result.
In particular cases, the DV auxiliary problem can be cast as that of finding the ground-state energy of a quantum system consisting of a particle in a potential well \cite{Touchette2018}.

However, the scaling \eqref{eq:DVscaling} has recently been observed to break down in numerous instances \cite{NT18, MeersonGaussian19, Jack20, Krajnik21}, where ``anomalous'' scalings were found.
Nickelsen and Touchette (NT) \cite{NT18} considered an Ornstein-Uhlenbeck (OU) process
\be
\label{eq:OU}
\dot{x}(t)=-\gamma x(t)+\sigma\eta (t) \, .
\ee
Here $x(t)$ is the position of the particle at time $t$, $\gamma > 0$ is the damping, $\eta(t)$ is white noise with $\left\langle \eta\left(t\right)\right\rangle =0$ and $\left\langle \eta\left(t\right)\eta\left(t'\right)\right\rangle =\delta\left(t-t'\right)$ where angular brackets denote ensemble averaging, and $\sigma > 0$ is the noise intensity. They studied the distribution of the observable \cite{footnote:IC}
\be
\label{eq:Adef}
A=\int_{0}^{T}x^{n} (t)dt \, .
\ee
where $n=1,2,\dots$. For $n=1$, $A$ can represent the work that laser tweezers perform when pulling on a Brownian particle \cite{zon2003a}, or the power dissipated in a noisy circuit \cite{zon2004a}. For $n=2,3$ and higher, $A$ is related to studies of fluctuations in turbulence velocity fields and small-scale intermittency \cite{matsumoto2013, nickelsen2017}.

NT probed the weak-noise limit $\sigma \to 0^+$ by employing the optimal fluctuation method (OFM) (sometimes also called the weak noise theory) \cite{Onsager,Freidlin,Dykman,Graham}.
They found that for $n>2$, the scaling \eqref{eq:DVscaling} breaks down and instead one observes an anomalous scaling
\be
\mathcal{P}\left(A;T, \sigma, \gamma \right)\sim\exp\left[-c_{n}\frac{\gamma^{\left(n+2\right)/n}A^{2/n}}{\sigma^{2}}\right],\quad \sigma \to 0^+,
\ee
where $c_{n}$ is an $n$-dependent constant that they calculated exactly. The physical mechanism behind this result is that the dominant contribution to $\mathcal{P}(A;T)$ comes from an instanton -- an optimal (most likely) trajectory $x(t)$. The instanton is the minimizer of the dynamical action $ \frac{1}{2}\int_{0}^{T}\left[\dot{x}\left(t\right)+x\left(t\right)\right]^{2}dt$ constrained on a given value of $A$, and it was found explicitly in \cite{MeersonGaussian19} (for completeness, we give the instanton in \eqref{eq:instanton}).
The instanton is (temporally) localized, that is, $x(t) \simeq 0$ at all times except for a short time window around $t=t_*$ for some $0 < t_* < T$.
The weak-noise limit $\sigma\to0^+$ corresponds to the far tail $|A| \to \infty$ of the distribution \cite{MeersonGaussian19}.
However, the full long-time distribution $\mathcal{P}\left(A;T\right)$ -- including typical fluctuations in addition to the distribution tails -- has been unknown, and its calculation is the main result of the current work.

\begin{figure*}[ht]
\includegraphics[width=0.485\linewidth,clip=]{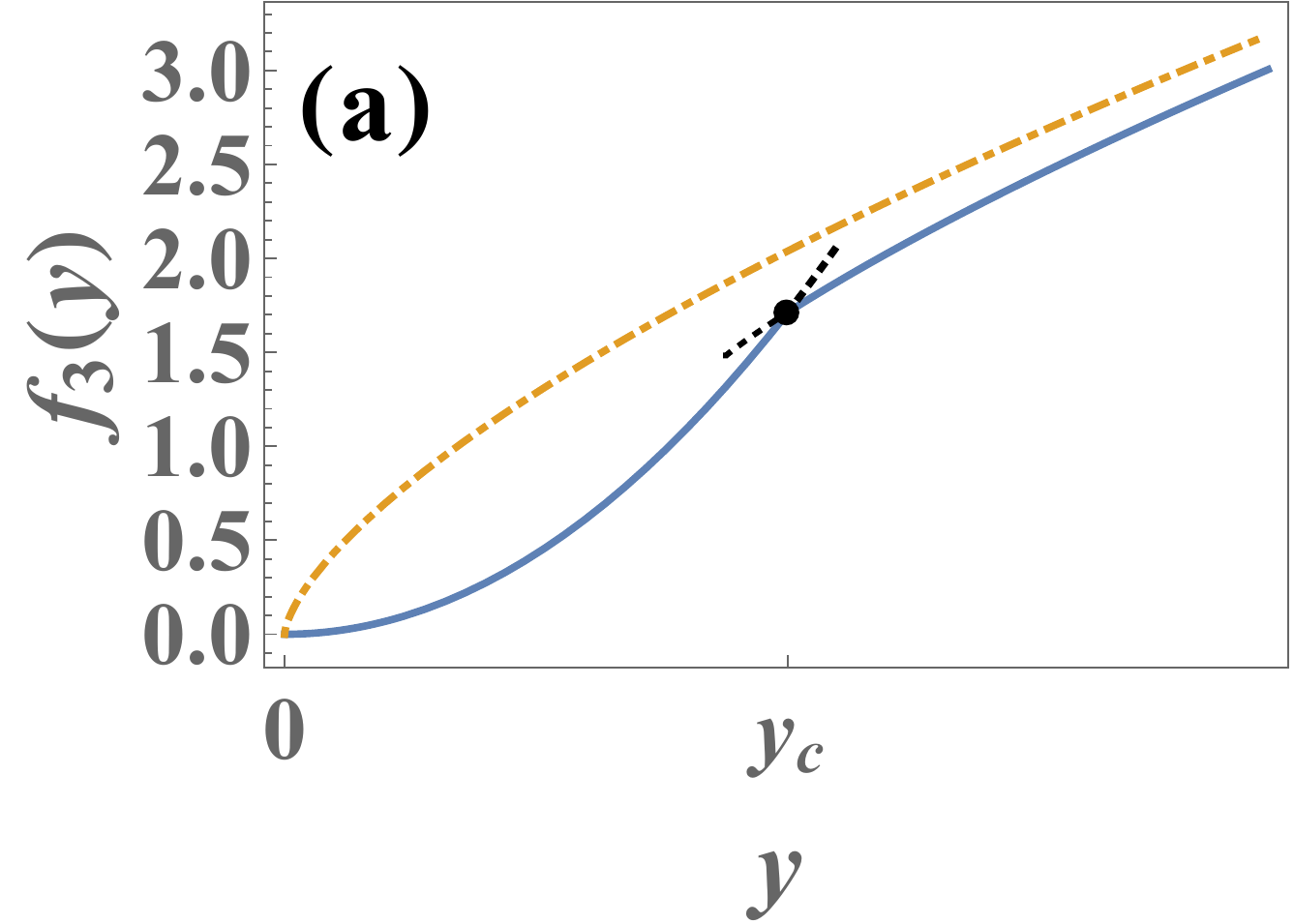}
\includegraphics[width=0.495\linewidth,clip=]{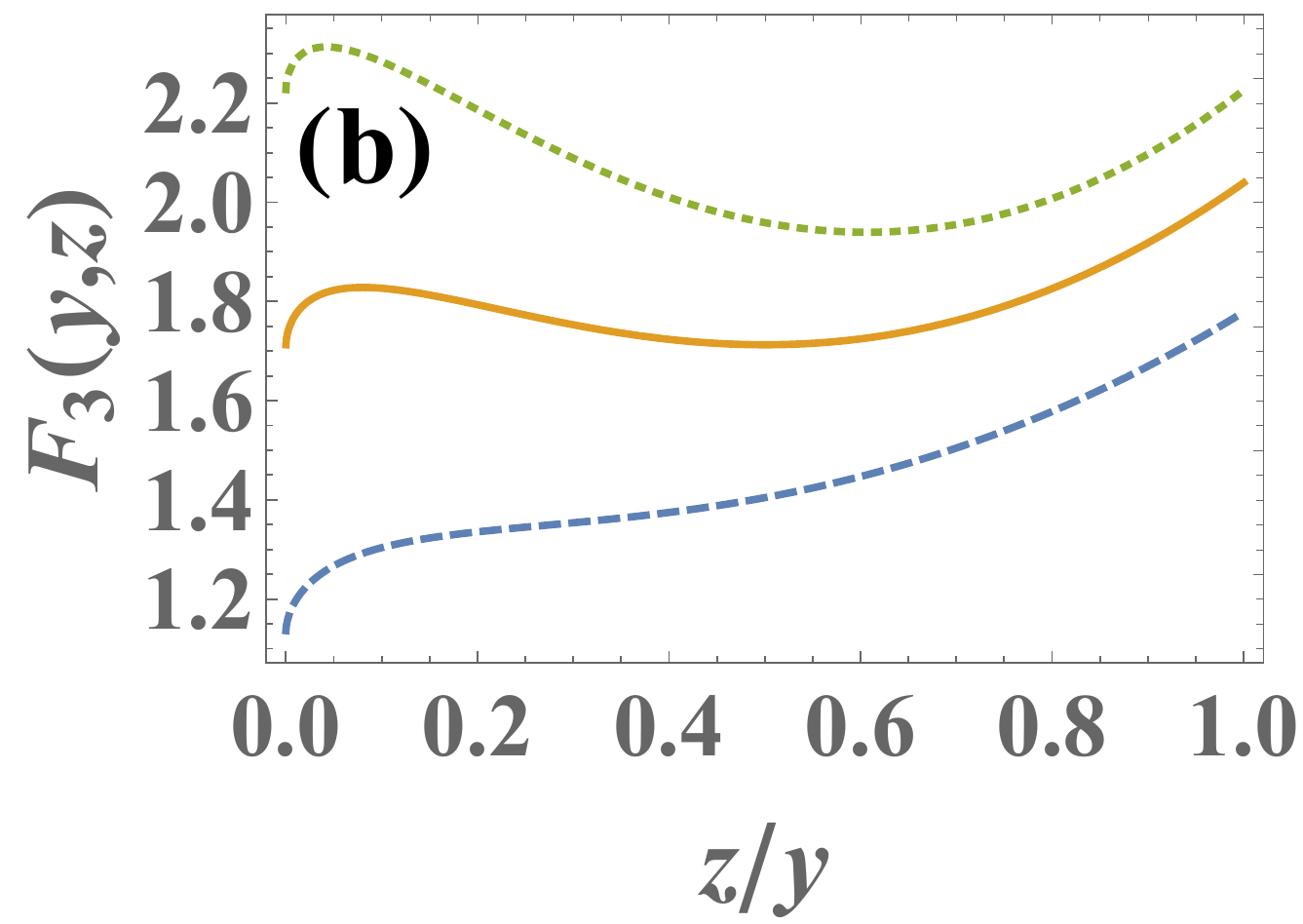}
\caption{(a) Solid line is the rate function $f_3(y)$ that describes the full distribution $P(A;T)$ for $n=3$, see Eqs.~\eqref{eq:PScaling}-\eqref{eq:Fndef}. Dotted lines are the continuations of the two branches of $f(y)$ into the regimes in which they are not optimal.
Dot-dashed line is the instanton's action $c_3 y^{2/3}$ which is strictly larger than $f(y)$, but gives the correct leading asymptotic behavior in the far tail of the distribution $y \gg 1$.
(b) $F_3(y,z)$ as a function of $z/y$ for three different values of $y$: subcritical $y = 5/2$ (dashed), critical $y = y_c$ (solid) and supercritical $y = 7/2$ (dotted). In the subcritical regime, the minimum of $F_3(y,z)$ is at $z=0$ whereas in the supercritical regime it is at $z=z_*(y) \ne 0$ given in Eq.~\eqref{eq:yfnyPara}.}
\label{fig:fandF}
\end{figure*}

We first point out that a similar anomaly can occur when considering distributions $P\left(S;N\right)$ of sums $S=\sum_{i=1}^{N}x_{i}$ of independent and identically distributed (i.i.d) random variables $x_1,\dots,x_N$, in the limit $N\gg1$. Here, the usual scaling is $P\left(S;N\right)\sim e^{-NI\left(S/N\right)}$ in analogy with \eqref{eq:DVscaling}. However, if the distribution tails of each of the $x_i$'s decays slower than exponentially, the usual scaling breaks down \cite{Nagaev69}, and in the far tail of $P\left(S;N\right)$ the ``big-jump'' principle holds: the dominant contribution to $P\left(S;N\right)$ comes from realizations where $S \simeq x_i$ for some $1\le i\le N$ \cite{Chistyakov,Foss,Denisov,Geluk,Clusel1}.
The big-jump principle has been observed in a wide range of systems 
including anomalous transport in  quenched disorder \cite{levyrand,Ub,VBB19} and
L\'evy walks \cite{VBB19,WVBB19,Gradenigo13,Barkai20}.
A ``condensation'' transition that separates between the typical-fluctuation regime, where the central limit theorem applies, and the distribution tail(s) where the big-jump principle applies, is a general phenomenon that has been observed in many instances \cite{EH05,M2008}. Examples include the zero-range process \cite{Grosskinsky03, Hirschberg15},
the discrete nonlinear
Schr\"odinger equation~\cite{RCK2000,SPP17,GIL21,GIL21b,GIP21}, economic and
financial models \cite{BM2000,BJJ2002,FZV13}, mass-transport
models
\cite{MEZ2005,EMZ06,EHM06,EM08,EMPT10,SEM2014,SEM2014b,SEM2016,GB2017},
and run-and-tumble active particles \cite{GM19,MoriGeneral21,Mori21}.
Above a critical point in the tail of $P\left(S;N\right)$, a condensate appears meaning that one of the $x_i$'s contributes a macroscopic fraction to $S$. In the far tail this fraction approaches unity, so the big-jump principle is recovered.
The condensation transition was shown to be universal for i.i.d random variables whose distribution decays slower than exponentially \cite{Brosset21, MoriGeneral21}.

We observe a striking similarity between the big jump in discrete-time systems and the instanton of NT in the continuous-time system \cite{NT18}. 
Both are localized events that dominate the contribution to the observable in question in the far distribution tail.
It is therefore appealing to search for a condensation transition in the distribution $\mathcal{P}(A;T)$. 
Indeed, since the stationary distribution of the OU process is Gaussian, we can gain intuition by considering an analogous discrete-time problem of the distributon of $\sum_{i=1}^{N}x_{i}^{n}$ where $x_{1},\dots,x_{N}$ are i.i.d. Gaussian random variables. This analogous problem exhibits a condensation transition, as can be shown using the general results of \cite{Brosset21, MoriGeneral21, Mori21}.
As we now show, such a transition is indeed present in our system too.

\section{Rescaling and summary of main results}
We begin by rescaling $\gamma t\to t$, $x\sqrt{\gamma}/\sigma\to x$, leading to the rescaled Langevin equation $\dot{x}=-x+\eta$ and to the (exact) scaling form
\be
\mathcal{P}\left(A;T,\sigma,\gamma\right)=\frac{\gamma^{\left(n+2\right)/2}}{\sigma^{n}}P\left(\frac{\gamma^{\left(n+2\right)/2}A}{\sigma^{n}};\gamma T\right)
\ee
of the distribution, where $P\left(A;T\right)$ is dimensionless and so are its (rescaled) arguments.
The weak-noise limit $\sigma \to 0^+$ is mathematically equivalent to the limit $|A| \to \infty$ in $P\left(A;T\right)$ \cite{MeersonGaussian19}. Therefore, the weak-noise results of \cite{NT18} describe the far tail(s) $|A| \to \infty$ of the distribution.

Let us state the main result of this paper. 
We study, for $n>2$, the distribution $P\left(A;T\right)$ in the long-time limit $T\gg1$. For $A > \left\langle A\right\rangle$ for even $n$, and all $A$ for odd $n$, we find that the distribution obeys a large-deviation principle with anomalous scaling exponents $\mu=2/\left(2n-2\right)$, and $\nu=n/\left(2n-2\right)$:
\be
\label{eq:PScaling}
P\left(A;T\right)\sim\exp\left[-T^{2/\left(2n-2\right)}f_{n}\left(\frac{\Delta A}{T^{n/\left(2n-2\right)}}\right)\right]
\ee
where $\Delta A = A - \left\langle A\right\rangle$,
\bea
\label{eq:fndef}
f_{n}\left(y\right)&=&\min_{z\in\left[0,y\right]}F_{n}\left(y,z\right),\\
\label{eq:Fndef}
F_{n}\left(y,z\right)&=&c_{n}z^{2/n}+\beta_{n}\left(y-z\right)^{2},
\eea
$c_n$ was calculated in \cite{NT18} (and for completeness, is given in Eq.~\eqref{eq:cnBn}) and we calculate $\beta_n$ [see Eq.~\eqref{eq:betan}] through a perturbative treatment of the DV auxiliary quantum problem, see Appendix \ref{app:DVPerturbation} for the details.
For instance, for $n=3$, $c_3=\left(9/10\right)^{1/3}$ \cite{NT18} and we find $\beta_3 = 2/11$, whereas for $n=4$, $c_4 = 2 / \sqrt{3}$ \cite{NT18} and we find $\left\langle A\right\rangle \simeq 3T/4$ and $\beta_4=2/21$.
The functions $f_3(y)$ and $F_3(y,z)$ are plotted in Fig.~\ref{fig:fandF}.
The result \eqref{eq:PScaling}-\eqref{eq:Fndef} is valid in the limit $T\to\infty$ with constant $\Delta A/T^{n/\left(2n-2\right)}$.
Remarkably, the first derivative of the rate function $f_n(y)$ has a discontinuity that can be interpreted as a first-order dynamical phase transition. For example, for $n=3$ this transition occurs at the critical values $y = \pm y_{c}$ where
$y_{c}\left(n=3\right)=\frac{11^{3/4}}{15^{1/4}}=3.069\dots$, and for $n=4$ the transition occurs at $y=y_{c}\left(n=4\right)=\frac{3^{4/3}7^{2/3}}{2^{5/3}}=4.987\dots$.
In the subcritical regime $\left|y\right|<y_{c}$ the rate function is purely parabolic $f_n\left(y\right)=\beta_{n}y^{2}$, describing a Gaussian distribution of typical fluctuations.
Interestingly, $f_n(y)$ is nonconvex.

As pointed out in \cite{NT18}, for even $n>2$, DV theory is valid at $0 < A < \left\langle A\right\rangle$ 
\cite{footnote:meanA}, so the scaling \eqref{eq:DVscaling} holds.
We now turn to the derivation of the results \eqref{eq:PScaling}-\eqref{eq:Fndef}. We treat only the anomalous case $n>2$, as for $n \le 2$ the scaling \eqref{eq:DVscaling} holds at all $A$, and $I(a)$ is found from DV theory \cite{NT18}.
Our strategy in the derivation is first to use a perturbative DV approach in order to treat the regime of typical fluctuations, and then to use this result in conjunction with the result for the tail $|A|\to\infty$ which is known from \cite{NT18}. Remarkably, this enables us to extract the entire intermediate regime $\Delta A\sim T^{n/\left(2n-2\right)}$. This is achieved by exploiting the separation of timescales between the duration of the instanton and the (much longer) duration $T$ of the entire dynamics.

\section{Typical fluctuations}
\begin{figure*}[ht]
\includegraphics[width=0.49\linewidth,clip=]{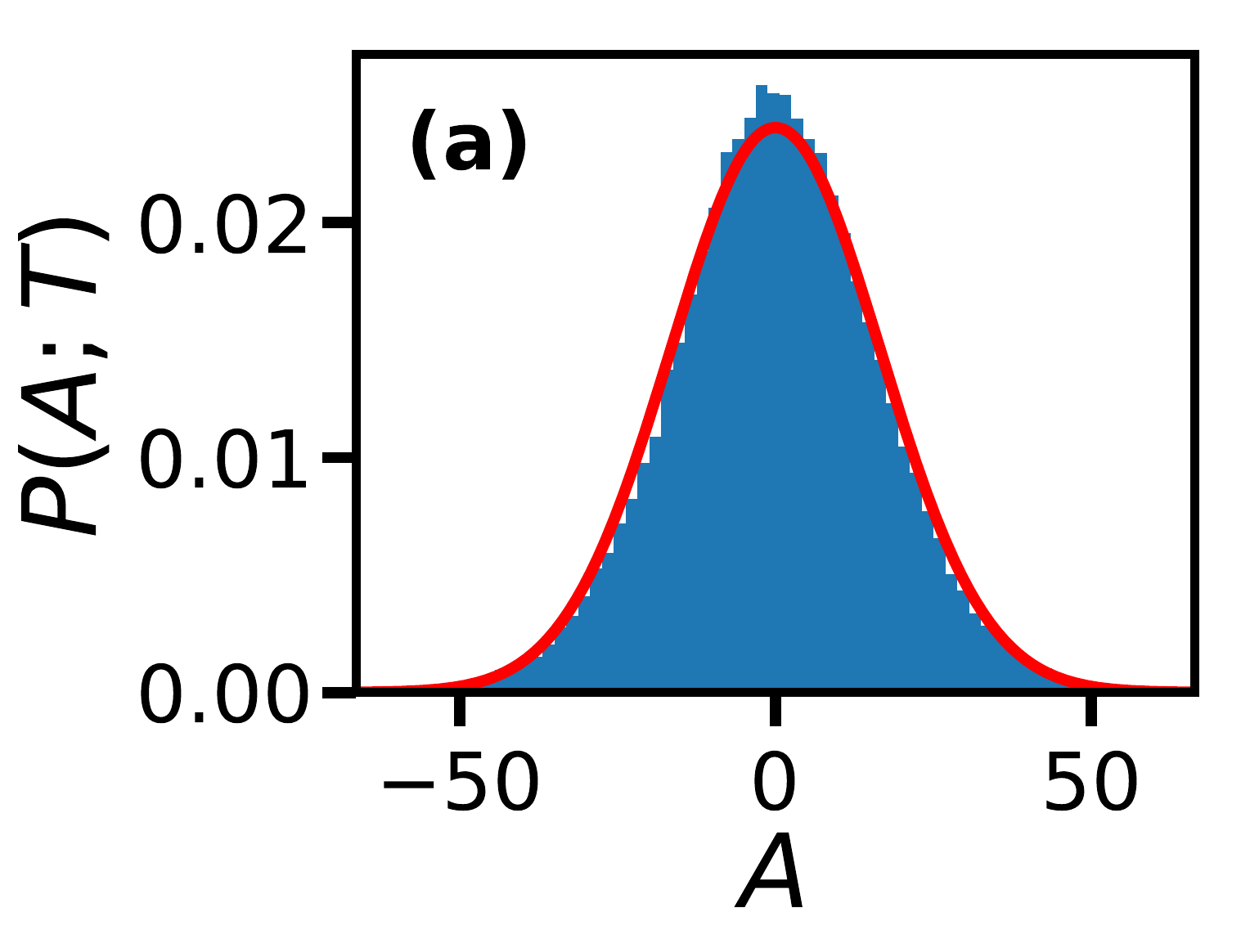}
\includegraphics[width=0.44\linewidth,clip=]{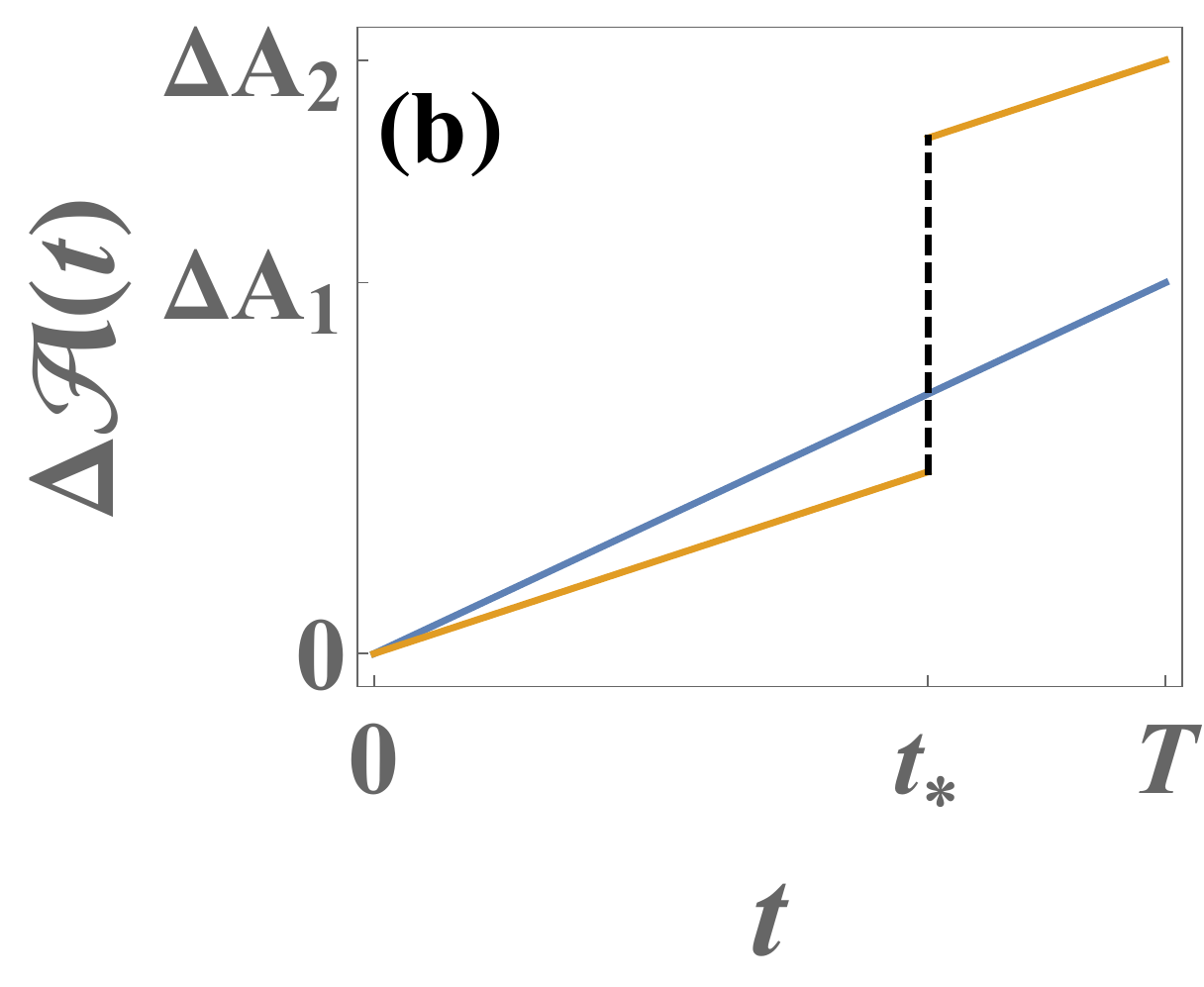}
\caption{(a) Bars represent a computation of $P(A;T)$ for $n=3$ and $T=100$ over ${10}^9$ Monte-Carlo simulations. Solid line is the Gaussian asymptotic \eqref{eq:Gaussian} with the normalization factor accounted for.
(b) An illustration of the optimal $\Delta \mathcal{A}(t)$ from \eqref{eq:calAoptimal} for two different values of $\Delta A$: subcritical $0 < \Delta A_1 < y_c T^{3/4}$ for which $\Delta A_{\text{ins}} = 0$, and supercritical  $\Delta A_2 > y_c T^{3/4}$ for which $\Delta A_{\text{ins}} \ne 0$. In the supercritical case, zooming in around the time $t_*$ where the instanton occurs, the discontinuity in \eqref{eq:calAoptimal} is smoothened out over a timescale of order unity, in which the corresponding trajectory $x(t)$ is described by the instanton \cite{NT18,MeersonGaussian19}.
}
\label{fig:AvstAndGauss}
\end{figure*}

We begin by considering the regime $|\Delta A| \ll T^{n/(2n-2)}$ which includes typical fluctuations.
The quantum potential whose ground-state energy one must calculate for a particle of unit mass (in units where $\hbar=1$) when using the DV method is
\be
\label{eq:Vkdef}
V_{k}\left(x\right)=\frac{x^{2}}{2}-\frac{1}{2}-kx^{n}
\ee
where $k$ is the tilt parameter and is related to the DV rate function $I(a)$ through a Legendre-Fenchel transform.
As pointed out in \cite{NT18}, the potential \eqref{eq:Vkdef} has no ground state due to the $x^n$ term (for $n>2$), signaling that the DV method breaks down.
However, we notice that at $|k| \ll 1$, the potential is effectively confining for $|x| \ll 1/|k|$, and its effective ground-state energy $ - \lambda\left(k\right)$ can be found perturbatively in $k$. 
Using second-order perturbation theory, we find (see Appendix \ref{app:DVPerturbation}) $\lambda\left(k\right)=\alpha_{n}k+k^{2}/4\beta_{n} + \dots$, where
\be
\label{eq:betan}
\alpha_{n}=\left\langle 0|x^{n}|0\right\rangle ,\quad\beta_{n}=\left(4\sum_{m=1}^{\infty}\frac{\left|\left\langle m|x^{n}|0\right\rangle \right|^{2}}{E_{m}-E_{0}}\right)^{-1}\!\!.
\ee
Here the $\left|m\right\rangle$'s are the energy eigenstates of the unperturbed ($k=0$) quantum oscillator with corresponding energies $E_m = m$.
The Legendre-Fenchel transform then yields the quadratic approximation of the DV rate function around its minimum, $I\left(a\right)=\beta_{n}\left(a-\alpha_{n}\right)^{2}+\dots$.
This predicts Gaussian fluctuations
\be
\label{eq:Gaussian}
P\left(A;T\right)\sim e^{-\beta_{n}\left(\Delta A\right)^{2}/T}
\ee 
around the mean value $\left\langle A\right\rangle\simeq\alpha_{n}T$ implying, in particular, that the variance of the distribution is $\text{Var}\left(A\right)\simeq T/2\beta_{n}$.
Note that $\alpha_n$ is simply the $n$th moment of the Gaussian distribution $P_{s}\left(x\right) =  e^{-x^{2}} / \sqrt{\pi}$, so $\alpha_{n}=2^{n/2}\Gamma\left(\frac{n+1}{2}\right)/\sqrt{\pi}$ in agreement with \cite{NT18}.
The prediction \eqref{eq:Gaussian} shows excellent agreement with a computation of $P(A;T)$ using Monte-Carlo simulations with $n=3$ and $T=100$, see Fig.~\ref{fig:AvstAndGauss}(a). 
The simulations were performed using an It\^{o} discretization of the (rescaled) Langevin equation \eqref{eq:OU} with time steps of size $0.01$.
Finally, the Legendre-Fenchel transform also yields the connection $a=\alpha_{n}+k/2\beta_{n} + \dots$.

We now analyze the regime of validity of the perturbative DV result \eqref{eq:Gaussian}. The last term in the tilted potential \eqref{eq:Vkdef} becomes of the same order as the first term at $x\sim1/\left|k\right|^{1/\left(n-2\right)}$.
One expects the DV calculation to be valid as long as the probability for the particle to reach this position is much smaller than $P\left(A;T\right)$ itself and therefore trajectories which reach such a position can be neglected when evaluating $P\left(A;T\right)$. Since the stationary distribution of the particle's position (for the OU process) is $P_{s}\left(x\right) =  e^{-x^{2}} / \sqrt{\pi}$, this validity condition yields
$1/\left|k\right|^{2/\left(n-2\right)}\gg \left(\Delta A\right)^{2}/T$
 which, using the connection $k \sim a - \alpha_n = \Delta A/T$, becomes $\left|\Delta A\right|\ll T^{n/\left(2n-2\right)}$.
As we will show shortly, we found that the perturbative DV result \eqref{eq:Gaussian} is actually valid in the entire subcritical regime $\left|\Delta A\right|/T^{n/\left(2n-2\right)} \le y_c$.
Finally, in this regime one easily checks that $|k| \ll 1$ (since $\left|\Delta A\right| \ll T$) , justifying the use of perturbation theory above.

\section{Full distribution}
\subsection{Mixed scenario}
We now have $P\left(A;T\right)$ at hand in two distinct regimes: in the far tail(s) $|A| \to \infty$, the result $P\left(A;T\right)\sim e^{-c_{n}A^{2/n}}$ of \cite{NT18}, and for typical fluctuations, the perturbative DV result \eqref{eq:Gaussian}. 
As we show below, the transition between the two regimes occurs at $\Delta A \sim T^{n/\left(2n-2\right)}$ where the two results predict probabilities of the same order.
In order to study the full distribution, it is therefore useful to take the scaling limit $T\to \infty$ with $\Delta A / T^{n/\left(2n-2\right)}$ constant.
The two regimes correspond to two different physical mechanisms for creating the large deviation. The instanton is a localized burst where the dominant contribution to the integral \eqref{eq:Adef} comes from a narrow temporal window. Defining 
\be
\Delta\mathcal{A}\left(t\right)=\int_{0}^{t}\left[x^{n}\left(s\right)-\left\langle x^{n}\left(s\right)\right\rangle \right]ds
\ee
[so $\Delta A=\Delta \mathcal{A}\left(T\right)$], the OFM (weak-noise) prediction corresponds to trajectories for which $\Delta \mathcal{A}\left(t\right)= \Delta A\theta\left(t-t_{*}\right)$ where $t_{*}\in\left[0,T\right]$ is the time when the instanton occurs and $\theta(t)$ is the Heaviside function.
To the contrary, the DV prediction corresponds to the opposite scenario, where the contribution to the integral \eqref{eq:Adef} is homogeneous throughout the entire dynamics because in the DV formalism, the conditioned process (i.e., the process $x(t)$ conditioned on a given value of $A$) becomes stationary in the large-$T$ limit. This leads to the linear behavior $\Delta \mathcal{A}\left(t\right)=\Delta At/T$.

The optimal scenario, however, can combine the two scenarios described above, as we now show. One part $\Delta A_\text{ins}$ of the integral \eqref{eq:Adef} could come from an instanton and the remainder, $\Delta A_{\text{DV}} = \Delta A-\Delta A_\text{ins}$, could come from the DV mechanism, corresponding to
\be
\label{eq:calAoptimal}
\Delta \mathcal{A}\left(t\right)=\Delta A_\text{ins}\theta\left(t-t_{*}\right)+\left(\Delta A-\Delta A_\text{ins}\right)t/T ,
\ee
see Fig.~\ref{fig:AvstAndGauss}(b).
Let us now calculate the probability of this mixed scenario.
A key observation is that, since the instanton is temporally localized while the DV mechanism is homogeneous over the entire duration of the dynamics,
the two mechanisms (instanton and DV) work independently from each other, so that the probability of this mixed scenario is 
\be
\label{eq:PAins}
\sim\exp\left[-c_{n}\Delta A_{\text{ins}}^{2/n}-\frac{\beta_{n}\left(\Delta A-\Delta A_{\text{ins}}\right)^{2}}{T}\right].
\ee
Importantly, for $\Delta A_{\text{ins}}\sim\Delta A\sim T^{n/\left(2n-2\right)}$, the two terms in the exponent in \eqref{eq:PAins} are of the same order.
Integrating the probability \eqref{eq:PAins} over $\Delta A_\text{ins}$ while using the saddle-point approximation (see Appendix \ref{sec:Fandf} for details), we obtain our main result reported above in Eqs.~\eqref{eq:PScaling}-\eqref{eq:Fndef}, describing a large-deviation principle with anomalous scaling and a rate function $f_n(y)$ which exhibits a first-order phase transition.
The optimal value of $\Delta A_{\text{ins}}$ in \eqref{eq:calAoptimal} is $\Delta A_{\text{ins}}=z \Delta A / y$ where $z$ is the minimizer in Eq.~\eqref{eq:fndef}.
%
In the subcritical regime $|y| < y_c$, the minimizer 
is at $z=0$, so $f_n\left(y\right)=\beta_{n}y^{2}$ is exactly parabolic, and the system is in a (temporally) homogeneous state.
In the supercritical regime $|y| > y_c$, the minimizer is at a nonzero value $z=z_*$, with the requirement $\partial F_n/\partial z=0$ giving $f_n(y)$ in a parametric form
\be
\label{eq:yfnyPara}
y=\frac{c_{n}}{n\beta_{n}}z_{*}^{\left(2-n\right)/n}+z_{*}, \; f_{n}\left(y\right)=c_{n}z_{*}^{2/n}+\frac{c_{n}^{2}}{n^{2}\beta_{n}}z_{*}^{2\left(2-n\right)/n},
\ee
and the system is in a ``condensed'' state.
For $n=3$, these equations can in fact be solved to find $f_n(y)$ explicitly but the result is very cumbersome so we do not give it here (a similar calculation was performed in \cite{Mori21}).
At $y=y_c$, $f_n(y)$ is continuous but its first derivative is not, see Fig.~\ref{fig:fandF} and Appendix \ref{sec:Fandf}.
The asymptotic behavior (obtained in Appendix \ref{sec:Fandf})
\be
\label{eq:fnasymptote}
f_{n}\left(\left|y\right|\gg1\right)\simeq c_{n}y^{2/n}-\frac{c_{n}^{2}}{n^{2}\beta_{n}}y^{2\left(2-n\right)/n}
\ee
describes the far tail(s) of the distribution,
the leading-order term coinciding with the OFM prediction of \cite{NT18}.

We now show that higher-order corrections to the DV perturbative result do not affect the result \eqref{eq:PScaling}-\eqref{eq:Fndef}. Higher-order perturbation theory in the DV formalism will produce correction terms of order $\left(\Delta A\right)^{m}/T^{m-1}$ with $m>2$ which will be added to Eq.~\eqref{eq:PScaling} in the exponent. However, when taking the limit $T\to \infty$ with $\Delta A/T^{n/\left(2n-2\right)}$ constant, these correction terms will have a vanishing contribution to $f_n,$ 
i.e., the correction terms do not affect the limit
$-\lim_{T\to\infty}\ln P\left(\left\langle A\right\rangle +yT^{n/\left(n-2\right)};T\right)/T^{2/\left(n-2\right)}=f_{n}\left(y\right)$.

\subsection{Optimality of the mixed scenario}
\label{sec:AN}
Strictly speaking, the mixed scenario described above gives, at this point, only an upper bound for the rate function $f_n(y)$ since we haven't yet shown that this scenario is optimal (i.e., that no other, likelier scenario exists). We now give a strong theoretical argument in favor of the optimality of the mixed scenario. Let us rewrite $A$ in the form $A=A_{1}+\dots+A_{N}$ where
\be
\label{eq:Ai}
A_{i}=\int_{\left(i-1\right)T/N}^{iT/N}x^{n}\left(t\right)dt,
\ee 
and $N$ is chosen such that $1 \ll N \ll T$. The $A_i$'s are correlated, but weakly so -- the correlation between $A_i$ and $A_j$ decays with $|i-j|$ over a characteristic scale that is much smaller than $N$.
It is natural to expect the general result of \cite{Brosset21}, obtained for i.i.d. random variables, to be valid for weakly correlated random variables too. Indeed, using that the variance of each of the $A_i$'s is $\text{Var}\left(A_{i}\right)\simeq\frac{T}{2\beta_{n}N}$ and that the tail of the PDF's of their distributions $P_i(A_i)$ behaves as $P_{i}\left(A_{i} \to \infty\right)\sim e^{-c_{n}A_{i}^{2/n}}$, our Eqs.~\eqref{eq:PScaling}-\eqref{eq:Fndef} are in perfect agreement with the calculation that uses the result of \cite{Brosset21}, see Appendix \ref{Ap:Brosset}.

\subsection{Conditioned process}

DV theory also predicts that the distribution $p\left(x|A\right)$ of $x = x(t)$ at some arbitrary intermediate time \cite{footnote:tIndependence}, conditioned on a given value of $A$, is given by the squared absolute value of the ground-state eigenfunction of the tilted potential \eqref{eq:Vkdef} \cite{Touchette2018}. As stated above, for our system the tilted potential has no ground state, but using the same logic as we used above, we obtain the effective ground state by treating $k$ perturbatively. This yields (see Appendix \ref{app:DVPerturbation} for details)
\bea
\label{eq:pxA}
p\left(x|A\right)&\simeq&P_s(x)-2k\sum_{m=1}^{\infty}\frac{\left\langle m|x^{n}|0\right\rangle }{E_{m}-E_{0}}\left\langle m|x\right\rangle \left\langle 0|x\right\rangle \nn\\
&+&\text{instanton contribution}
\eea
%
where, using the connection between $a$ and $k$ found above, $k \! = \! 2\beta_{n}\Delta A_{\text{DV}}/T$.
Due to the localization of the instanton, the last term in Eq.~\eqref{eq:pxA} is of order $1/T \! \ll \! 1$. 
For $n \! = \! 3$ Eq.~\eqref{eq:pxA} reads $p\left(x|A\right)\simeq\left[1-k\left(\frac{2}{3}x^{3}+2x\right)\right]e^{-x^{2}} / \sqrt{\pi}+\text{instanton contribution}$ (the calculation is performed explicitly in Appendix \ref{app:DVPerturbation}).

\section{Summary and discussion}
We calculated the distribution $P(A;T)$ for $n>2$ in the long-time limit $T\gg1$ for $A > \left\langle A\right\rangle$ for even $n$, and all $A$ for odd $n$, see Eqs.~\eqref{eq:PScaling}-\eqref{eq:Fndef}.
We showed that the two main generic tools in the study of large deviations, namely DV and the OFM, correspond to scenarios that describe the fluctuations of $A$ in different regimes. We uncovered a remarkable first-order dynamical phase transition, corresponding to a jump in the first derivative of the rate function $f_n(y)$. In the subcritical regime the fluctuations of $A$ are Gaussian, while in
the supercritical regime the optimal scenario leading to a given value of $A$ is a combination of the DV and instanton scenarios, described by \eqref{eq:calAoptimal}, that dominates the contribution to $P(A;T)$.
We also calculated the conditional distribution of $x(t)$ at an intermediate time, conditioned on a given value of $A$.

For even $n$, at $0 < A < \left\langle A\right\rangle$, $P(A;T)$ is described by Eq.~\eqref{eq:DVscaling} where $I(a)$ is found from DV theory \cite{NT18}. At $A \simeq \left\langle A\right\rangle$ the two results \eqref{eq:DVscaling} and \eqref{eq:PScaling} match smoothly due to the parabolic behaviors (with the same coefficient $\beta_n$) of the rate functions $f_n(y)$ and $I(a)$ around their minima.

It is worth noting that our anomalous exponents, $\mu=2/\left(2n-2\right)$ and $\nu=n/\left(2n-2\right)$, are different to those found by NT \cite{NT18}. In contrast to NT's result which is valid only in the very far tail,
the rate function $f_n(y)$ given in the present work describes the entire distribution $P(A;T)$: from typical Gaussian fluctuations up to the far tail that is dominated by the instanton in the leading order.

It would be interesting to observe the regime of the transition in numerical simulations. As shown in Fig.~\ref{fig:AvstAndGauss}(a), the theory shows good agreement with the simulations in the typical-fluctuations regime. The far-tail result showed good agreement with simulations in \cite{NT18}, which were performed with $T=30$. However, capturing the transition regime is far more challenging: We found that $T=100$ was not large enough to observe the convergence to \eqref{eq:PScaling} in the transition regime (not shown), so longer simulation times and special sampling methods \cite{Hartmann2002} are needed in order to observe this regime.

It would be interesting to explore a possible universality of the condensation transition found here in a broader class of continuous-time systems, in analogy with that found in \cite{Brosset21, MoriGeneral21, Mori21} 
for the distribution of sums of $N$ i.i.d. random variables.
Remarkably, our rate function $f_n(y)$ coincides, up to the constants $c_n$ and $\beta_n$, with those found in \cite{Brosset21, MoriGeneral21, Mori21}. This connection may be related to the argument given in section \ref{sec:AN}, formulating our problem in terms of the distribution of a sum of weakly correlated random variables.
Our perturbative DV approach appears to be a general method for calculating the variance of typical, Gaussian fluctuations in a broad class of continuous-time systems even when the DV tilted operator has no ground state. In fact, the perturbative approach is well-known in the context of quantum mechanics, as it describes long-lived bound states with a complex energy describing a decay \cite{Vainshtein64, Baz71}. Similarly, the OFM appears to be a general method for calculating the far tail(s) in such systems, and the two methods together give the full rate function.

In particular, one could search for a similar transition in a class of Gaussian, but not necessarily Markovian, processes (which includes the OU process as a particular case) studied in \cite{MeersonGaussian19}, whose fluctuations were also shown to exhibit anomalous scaling.
For these processes the anomalous scaling $P\left(A;T\right)\sim e^{-T^{\mu}f\left(\Delta A/T^{\nu}\right)}$ was conjectured and it was shown that from the behavior in the far tail, one can deduce the relation $\nu = n \mu /2$ \cite{MeersonGaussian19} which is indeed satisfied by our anomalous exponents in \eqref{eq:PScaling}.

\textit{Acknowledgments}
I warmly acknowledge helpful discussions with Tal Agranov, Hugo Touchette, Satya N. Majumdar, Alex Kamenev and Baruch Meerson, and I thank the latter for a critical reading of the manuscript.
I acknowledge support from the Yad Hanadiv fund (Rothschild fellowship).

\appendix

\section{Obtaining $F_n(y,z)$ and $f_n(y)$ and deriving some of their properties}
\label{sec:Fandf}

In order to obtain our results \eqref{eq:PScaling}-\eqref{eq:Fndef} from \eqref{eq:PAins} (all in the main text), we first rewrite 
the latter equation in the form $e^{-S\left(\Delta A_{\text{ins}}\right)}$ where 
\bea
\label{eq:Sdef}
&&S\left(\Delta A_{\text{ins}}\right)=c_{n}\Delta A_{\text{ins}}^{2/n}+\beta_{n}\frac{\left(\Delta A-\Delta A_{\text{ins}}\right)^{2}}{T} \nn\\
&&\qquad =T^{2/\left(2n-2\right)}F_{n}\left(\frac{\Delta A}{T^{n/\left(2n-2\right)}},\frac{\Delta A_{\text{ins}}}{T^{n/\left(2n-2\right)}}\right).
\eea
$P\left(A;T\right)$ is now obtained by integrating 
\bea
\label{eq:SPA}
P\left(A;T\right)&\sim&\int_{0}^{\Delta A}e^{-S\left(\Delta A_{\text{ins}}\right)}d\Delta A_{\text{ins}} \nn\\
&\sim&\int_{0}^{y}e^{-T^{2/\left(2n-2\right)}F_{n}\left(y,z\right)}dz
\eea
where we are considering here the large-$T$ limit with $y=\Delta A/T^{n/\left(2n-2\right)}$ constant (because in this limit that the two terms in $S$ are of the same order).
In the long-time limit, the large parameter $T^{2/\left(2n-2\right)} \gg 1$ in the exponent ensures that the dominant contribution to the integral \eqref{eq:SPA} comes from saddle-point approximation, $P\left(A;T\right)\sim e^{-T^{2/\left(2n-2\right)}F_{n}\left(y,z_{*}\right)}$ where $z_*$ is the minimizer of $F_n(y,z)$ over $z$, which is precisely Eqs.~\eqref{eq:PScaling}-\eqref{eq:Fndef} in the main text.

We now analyze properties of the functions $F_n(y,z)$ and $f_n(y)$, and in particular we focus on the first-order dynamical phase transition in $f_n(y)$.
For odd $n$, the distribution is clearly symmetric $P\left(A;T\right)=P\left(-A;T\right)$ so that $f_n(y) = f_n(-y)$, whereas for even $n$, the scaling form \eqref{eq:PScaling} in the main text is only valid at $y>0$. Therefore, for simplicity, let us assume here that $y>0$.
%
%
The local minimum of $F_n(y,z)$ at the nonzero value $z=z_*$ only exists at $y > y_\ell$ where 
\be
\left.\frac{\partial^{2}F_{n}\left(y_{\ell},z\right)}{\partial z^{2}}\right|_{z=z_{\ell}\equiv z_{*}\left(y_{\ell}\right)}=0
\ee
leading to 
\be
y_{\ell}=\frac{2n-2}{n-2}\left[\frac{n^{2}\beta_{n}}{\left(n-2\right)c_{n}}\right]^{n/\left(2-2n\right)}\,.
\ee
This local minimum becomes the \emph{global} minimum at $y>y_{c}$ which we find from the continuity of $f_n(y)$ at $y=y_{c}$,
\be
y_c=\frac{c_{3}}{3\beta_{3}}z_{c}^{-1/3}+z_{c},\quad\beta_{3}y_{c}^{2}=c_{3}z_{c}^{2/3}+\frac{c_{3}^{2}}{9\beta_{3}}z_{c}^{-2/3}.
\ee
We find 
\bea
z_{c}&=&\left[\frac{\left(n-2\right)c_{n}}{n\beta_{n}}\right]^{n/\left(2n-2\right)},\\
\label{eq:zcyc}
y_{c}&=&\frac{n-1}{n-2}\left[\frac{\left(n-2\right)c_{n}}{n\beta_{n}}\right]^{n/\left(2n-2\right)}.
\eea
For $n\in\left\{ 3,4\right\} $ this gives the critical values
\bea
&&\!\!\!\! y_{c}\left(n=3\right)=2\left(\frac{c_{3}}{3\beta_{3}}\right)^{3/4}=\frac{11^{3/4}}{15^{1/4}}=3.069\dots,\\
&&\!\!\!\! y_{c}\left(n=4\right)=\frac{3}{2^{5/3}}\left(\frac{c_{4}}{\beta_{4}}\right)^{2/3} \!\!=\frac{3^{4/3}7^{2/3}}{2^{5/3}}=4.987\dots.
\eea
At $y=y_c$,  $f_n(y)$ is continuous but its first derivative jumps. Indeed, $f_{n}'\left(y_{c}^{-}\right)=2\beta_{n}y_{c}$ while, in the supercritical regime one finds from the chain rule
\be
\frac{df_{n}}{dy}=\frac{df_{n}/dz_{*}}{dy/dz_{*}}=\frac{2c_{n}}{n}z_{*}^{\left(2-n\right)/n}
\ee
so $f_{n}'\left(y_{c}^{+}\right)=\left(2c_{n}/n\right)z_{c}^{\left(2-n\right)/n}$. Using Eq.~\eqref{eq:zcyc} one finds $f_{n}'\left(y_{c}^{-}\right)=\left(n-1\right)f_{n}'\left(y_{c}^{+}\right)$ so the two one-sided derivatives indeed differ.
Finally, at $y\gg1$, Eq.~\eqref{eq:yfnyPara} yields $z_{*}\simeq y-\frac{c_{n}}{n\beta_{n}}y^{\left(2-n\right)/n}$ leading to the asymptotic behavior
\be
f_n\left(y\gg1\right)\simeq c_{n}y^{2/n}-\frac{c_{n}^{2}}{n^{2}\beta_{n}}y^{2\left(2-n\right)/n}
\ee
given also in Eq.~\eqref{eq:fnasymptote} of the main text.

  	
\section{Perturbation theory on the DV quantum potential}
\label{app:DVPerturbation}

The Donsker-Varadhan formalism \cite{DonskerVaradhan} reduces the problem of finding the rate function $I(a)$ from Eq.~\eqref{eq:DVscaling} of the main text, to that of finding the largest eigenvalue $\lambda(k)$ of the ``tilted'' operator $\frac{1}{2}\partial_{x}^{2}-V_{k}\left(x\right)$. Equivalently, $-\lambda(k)$ is the ground-state energy for a quantum particle of unit mass in the ``tilted'' potential $V_{k}\left(x\right)$ (in units where $\hbar=1$).
According to the G\"{a}rtner-Ellis theorem \cite{Ellis}, the DV rate function $I(a)$ is recovered from $\lambda(k)$ by applying a Legendre-Fenchel transform
	\begin{equation}\label{GE}
	I(a) = \sup_{k \in \mathbb{R}} \left[ k a - \lambda(k) \right] \, .
	\end{equation}
However, for the Ornstein-Uhlenbeck process and $A$ defined as in Eq.~\eqref{eq:Adef} in the main text, the potential is given by
\be
V_{k}\left(x\right)=\frac{x^{2}}{2}-\frac{1}{2}-kx^{n}.
\ee
As pointed out in \cite{NT18}, for $n>2$, this potential clearly has no ground state for $k \ne 0$, signaling that DV breaks down here.
Still, following the argument given in the main text, we can calculate the effective ground-state energy perturbatively in the parameter $|k| \ll 1$, and we expect the result to correctly describe typical fluctuations of $A$.

At $k=0$ the ground-state energy vanishes and the corresponding wave function is $\left\langle 0|x\right\rangle = \pi^{-1/4} e^{-x^{2}/2}$.
Second-order pertubation theory yields
\be
\label{eq:lambdak}
\lambda\left(k\right)=k\left\langle 0|x^{n}|0\right\rangle +k^{2}\sum_{m=1}^{\infty}\frac{\left|\left\langle m|x^{n}|0\right\rangle \right|^{2}}{E_{m}-E_{0}} + \dots
\ee
where $E_m = m$ are the unperturbed energy levels.
The matrix elements $\left\langle m|x^{n}|0\right\rangle $ are straightforward to calculate by expressing $x=\left(a+a^{\dagger}\right)/\sqrt{2}$ using creation and annihilation operators, and one finds that the sum in \eqref{eq:lambdak} includes only a finite number of nonzero terms.
For $n=3 $ one finds, using 
$\left(a+a^{\dagger}\right)^{3}\left|0\right\rangle =3\left|1\right\rangle +\sqrt{6}\left|3\right\rangle$,
that  $\lambda\left(k\right)=11k^{2}/8 + \dots$, and the Legendre-Fenchel transform \eqref{GE} then gives the connection $a=\lambda'\left(k\right)=11k/4+ \dots$ leading to $I\left(a\right)=2a^{2}/11 + \dots$, so $\beta_3 = 2/11$.
Similarly, for $n=4$, $\lambda\left(k\right)=3k/4+21k^{2}/8+ \dots$, $a\left(k\right)=3/4+21k/4+ \dots$ and $I\left(a\right)=2\left(a-3/4\right)^{2}/21 + \dots$, so $\beta_4 = 2/21$.
Here the minimum of $I(a)$ is at $a=3/4$ which corresponds to the mean value $\left\langle A\right\rangle \simeq 3T/4$, and it equals the value predicted by the stationary distribution $\left\langle A\right\rangle = T\int_{-\infty}^{\infty}x^{4}P_{s}\left(x\right)dx$.
For general $n$ one has
\be
I(a) = \beta_{n}\left(a-\left\langle 0|x^{n}|0\right\rangle \right)^{2},\quad\beta_{n}=\left(4\sum_{m=1}^{\infty}\frac{\left|\left\langle m|x^{n}|0\right\rangle \right|^{2}}{E_{m}-E_{0}}\right)^{-1}
\ee
which is essentially equivalent to Eq.~\eqref{eq:betan} of the main text.
Finally, the ground state, including its first-order perturbative correction, is given by 
\be
\left|\psi\right\rangle =\left|0\right\rangle -k\sum_{m=1}^{\infty}\frac{\left\langle m|x^{n}|0\right\rangle }{E_{m}-E_{0}}\left|m\right\rangle + \dots .
\ee
The conditional distribution $p(x|A)$ defined in the main text is then given by $\left\langle x|\psi\right\rangle ^{2}$, which, using $P_{s}\left(x\right)=\left|\left\langle 0|x\right\rangle \right|^{2}$ and keeping the leading-order term in $k$, yields Eq.~\eqref{eq:pxA} of the main text without the last term on the right hand side (which is not accounted for by DV theory).
For $n=3$, using $\left(a+a^{\dagger}\right)^{3}\left|0\right\rangle =3\left|1\right\rangle +\sqrt{6}\left|3\right\rangle$ and plugging the wave functions of the (unperturbed) harmonic oscillator
\bea
\left\langle 0|x\right\rangle &=&\pi^{-1/4}e^{-x^{2}/2},\\
\left\langle 1|x\right\rangle &=&\pi^{-1/4}\sqrt{2}\,xe^{-x^{2}/2},\\
\left\langle 3|x\right\rangle &=&\pi^{-1/4}\frac{8x^{3}-12x}{\sqrt{48}}e^{-x^{2}/2}
\eea
into Eq.~\eqref{eq:pxA} of the main text, we obtain the result reported just below it.

\section{Additional details regarding the instanton}

In \cite{NT18} the instanton's action
\be
\label{eq:actioncn}
\frac{1}{2}\int_{0}^{T}\left[\dot{x}\left(t\right)+x\left(t\right)\right]^{2}dt=c_{n}\left(\Delta A_\text{ins}\right)^{2/n}
\ee
was calculated. The expression for $c_n$ given there can in fact be simplified a little and written as 
\be
\label{eq:cnBn}
c_{n}=\frac{nB_{n}}{8}, \quad B_{n}=2\left[\frac{2\sqrt{\pi}\,\Gamma\left(\frac{n}{n-2}\right)}{(n-2)\Gamma\left(\frac{3n-2}{2\left(n-2\right)}\right)}\right]^{\left(n-2\right)/n}
\ee
where $\Gamma\left(z\right)=\int_{0}^{\infty}t^{z-1}e^{-t}dt$ is the gamma function.
In \cite{MeersonGaussian19} the instanton itself was also found:
\bea
\label{eq:instanton}
x\left(t\right)&=&\left(2B_{n}\right)^{1/\left(2-n\right)}\left(\Delta A_\text{ins}\right)^{1/n} \nn\\
&\times& \left[\text{sech}\left(\frac{\left(n-2\right)\left(t-t_{*}\right)}{2}\right)\right]^{2/\left(n-2\right)}\,.
\eea
As described in the main text, one indeed finds that the instanton is (temporally) localized, that is, $x(t) \simeq 0$ except for a short time window (whose duration, of order unity, is much shorter than $T$) around the time $t_*$.
Note that Eqs.~\eqref{eq:actioncn} and \eqref{eq:instanton} were originally given (in \cite{NT18} and \cite{MeersonGaussian19} respectively) with $\Delta A_\text{ins}$ replaced by $A$.
The reason for the difference is that they worked in the low noise limit $\sigma \to 0^+$ in the physical variables, which, as shown in the present work, is equivalent to the far distribution tail(s) $|A| \to \infty$. In their limit $|A| \gg \left\langle A\right\rangle$ and, as argued in the main text, the instanton dominates the contribution to $\Delta A$, so $A \simeq \Delta A \simeq \Delta A_\text{ins}$. In this work, however, here we work in the scaling limit $\Delta A\sim T^{n/\left(2n-2\right)}$ and thus it is important to leave $\Delta A_\text{ins}$ as it is in Eqs.~\eqref{eq:actioncn} and \eqref{eq:instanton}.

\section{Comparison with Brosset et. al.}
\label{Ap:Brosset}
In Ref.~\cite{Brosset21}, distributions of sums $S=X_1 + \dots + X_N$ of i.i.d. random variables were studied. Assuming that each of the $X_i$'s has zero mean and variance $V$ and that the distribution tails are stretched exponentials,
\be
\ln\mathbb{P}\left(X\ge x\right)\sim-qx^{1-\epsilon} \, .
\ee
They proved that for all $C>0$,
\bea
&&\lim_{N\to\infty}\frac{N}{C^{2}N^{2/\left(1+\epsilon\right)}}\ln\mathbb{P}\left(S\ge CN^{1/\left(1+\epsilon\right)}\right) \nn\\
&&\quad\quad =-\inf_{0\le t\le1}\left\{ \frac{q\left(1-t\right)^{1-\epsilon}}{C^{1+\epsilon}}+\frac{t^{2}}{2V}\right\}  \, .
\eea
In order to compare with our results, it is convenient to re-write this in the form
\bea
\label{eq:PBrosset}
&&\mathbb{P}\left(\frac{S}{N^{1/\left(1+\epsilon\right)}}\ge C\right)\sim \nn\\
&&\exp\left[-\frac{C^{2}N^{2/\left(1+\epsilon\right)}}{N}\inf_{0\le t\le1}\left\{ \frac{q\left(1-t\right)^{1-\epsilon}}{C^{1+\epsilon}}+\frac{t^{2}}{2V}\right\} \right]
\eea
which is valid in the limit $N \gg 1$.

Using the argument in section \ref{sec:AN}, we now show that our Eqs.~\eqref{eq:PScaling}-\eqref{eq:Fndef} are in perfect agreement with \cite{Brosset21}.
For convenience, let us first consider odd $n$, for which the means of the $A_i$'s vanish.
Replacing $S\to A$, $X_i \to A_i$ [defined in Eq.~\eqref{eq:Ai}], $V \to \frac{T}{2\beta_{n}N}$, $1-\epsilon \to 2/n$ and $q \to c_n$,
and replacing the cumulative distribution functions by the PDFs (because, in the large-deviation regime they are equal up to a subleading prefactor), Eq.~\eqref{eq:PBrosset} becomes
\be
P\left(A\right) \sim \exp\left[-\inf_{0\le t\le1}\left\{ c_{n}\left(1-t\right)^{1-\epsilon}A^{1-\epsilon}+\frac{\beta_{n}t^{2}A^{2}}{T}\right\} \right]
\ee
which indeed coincides with our Eqs.~\eqref{eq:PScaling}-\eqref{eq:Fndef}.
For even $n$, the same argument with the replacement $S=\Delta A$, $X_{i}=A_{i}-\left\langle A_{i}\right\rangle$ works.


\begin{thebibliography} {99}
\bibitem{footnote:IC} To complete the defintion of the problem, one should specify an initial condition $x(t=0) = x_0$. However, in the long-time limit that we are considering, the initial condition becomes unimportant.
\bibitem{footnote:meanA} $\left\langle A\right\rangle$ is easy to calculate from the stationary distribution, $\left\langle A\right\rangle \simeq T\int_{-\infty}^{\infty}x^{n}P_{s}\left(x\right)dx$, using the fact that the system is ergodic and therefore time-averaged quantities converge, for large times, to their ensemble-averaged values.
\end{thebibliography}
\end{document}